\documentclass[prl,aps,twocolumn,showpacs,amsmath,amssymb]{revtex4}

\usepackage{graphicx}
\usepackage{color}
\usepackage{bbm} 
\usepackage{dsfont}

\begin{document}

\title{Confinement limit of Dirac particles in scalar 1D potentials}
\author{R. G. Unanyan$^{1,2}$, J. Otterbach$^1$, and M. Fleischhauer$^1$}
\affiliation{$^{1}$Department of Physics and Research Center OPTIMAS, Technische Universit\"{a}t Kaiserslautern, 67663,
Kaiserslautern, Germany}
\affiliation{$^{2}$Institute for Physical Research Armenian National Academy of Sciences,
Ashtarak-2 378410, Armenia}
\begin{abstract}
We present a general proof that Dirac particles cannot be localized below their Compton length by symmetric but otherwise arbitrary scalar potentials. This proof does not invoke the Heisenberg uncertainty relation and thus does not rely on the nonrelativistic linear momentum relation. Further it is argued that the result is also applicable for more general potentials, as e.g. generated by nonlinear interactions. Finally a possible realisation of such a system is proposed.
\end{abstract}


\maketitle

\section{Introduction}
It is generally accepted that relativistic particles with rest mass $m_0$ described by the Dirac equation cannot be localized better than half its corresponding Compton wave length $\lambda_C\equiv\hbar/m_0c^2$ \cite{Bjorken}. The argument to support this fact is based on the Heisenberg uncertainty relation $\Delta z\Delta p\geq\frac{\hbar}{2}$. Using the relation $\Delta p=m_0\Delta v$ and observing that $\Delta v<c$ one arrives at the statement $\Delta z\geq\hbar/2m_0c^2=\lambda_C/2$. However this argument is not stringent since it is based on the purly nonrelativistic relation $p=m_0v$ and can therefore not be used for highly relativistic particles. 

Recently there is a growing interest in various realisations of systems exposing Dirac-like features \cite{graphene,atoms,merkl-EPL-2008}. In \cite{Otterbach-PRL-2009} it is shown that stationary light \cite{Zimmer-OptComm-2006} at small spatial length scales exhibits Dirac-like behavior, as e.g. Zitterbewegung and Klein tunneling.It is in this context that the question about the lower confinement limit of Dirac particles arises.

For freely moving Dirac particles a stringent derivation of the lower localization bound can be found in \cite{dodonov-PLA}

In this brief report we want to provide a general proof of the statement that the coordinate variance $\langle\Delta z\rangle$ for any Dirac particle with rest mass $m_0$ in a one dimensional symmetric scalar potential is strictly larger than half its corresponding Compton length. In doing so we neither use the Heisenberg uncertainty relation nor the nonrelativistic linear momentum relation. We also argue that the result can be generalized to potentials created by certain nonlinear interactions. A well known example for this kind of interactions is the massive Thirring model \cite{Thirring}.

\section{Proof}
The one dimensional Dirac equation for a particle with rest mass $m_0$ in external scalar and vector potentials $U(z)$, $A(z)$ is given by
\begin{equation}
 i\hbar\partial_{t}\Psi(t,z)=\big[c\alpha\left(-i\hbar \partial_z-A(z)\right)+\beta m_0c^2+U(z)\big]\Psi(t,z).
\label{eq:Dirac}
\end{equation}

It can be shown \cite{Martino-PRL-2007} that the case of a vanishing scalar potential, i.e. $U(z)=0$, eq. (\ref{eq:Dirac}) is equivalent to an effective Schr\"odinger equation. Thus confinement to arbitrarily small length scales is possible.

In the following we assume that $A(z)=0$ and only a finite scalar potential $U(z)$ is present. As a particular representation of the Dirac algebra we choose
\begin{equation}
 \alpha=\begin{pmatrix}
         0 & -i \\ i & 0
        \end{pmatrix}
\quad
 \beta=\begin{pmatrix}
        1 & 0 \\ 0 & -1
       \end{pmatrix}.
\end{equation}
This is a permissible choice, since it fulfills the restrictions $\alpha^2=\beta^2=\mathds{1}_{2\times 2}$, $\alpha\beta+\beta\alpha=0$ and leaves eq.(\ref{eq:Dirac}) invariant under Lorentz transformations \cite{Scheck}. Hence, in 1D it is sufficient to use a single spinor with two scalar components $\Psi(t,z)=(\phi(t,z),\chi(t,z))^T$. Making the ansatz $\Psi(t,z)=\Psi(z)\exp[-iEt/\hbar]$ and introducing dimensionless quantities
\begin{equation}
 \gamma=\frac{E}{m_0c^2}\quad z\rightarrow\frac{z}{\lambda_C},\quad \lambda_C=\frac{\hbar}{m_0c},
\label{eq:dimensionlessQuantities}
\end{equation}
where $\lambda_C$ is the Compton length, we arrive at the following coupled differential equations
\begin{align}
 \gamma \phi(z)=&-\frac{\partial}{\partial z}\chi(z)+\big(1+f(z)\big)\phi(z) \\
 \gamma \chi(z)=&\frac{\partial}{\partial z}\phi(z)-\big(1-f(z)\big)\chi(z).
\label{eq:DiracDimensionless}
\end{align}
Here $f(z)=U(z)/m_0c^2$ is the normalized external potential.

In the following we want to derive an estimation for the coordinate variance $\langle\Delta z\rangle=\sqrt{\langle z^2\rangle-\langle z\rangle^2}$. Expectation values of any Operator $A$ are calculated by
\begin{equation}
 \langle A\rangle=\int\limits_{-\infty}^\infty\text{d}z \Psi^\dagger (z)\,A\, \Psi(z).
\end{equation}

First we calculate the expectation value of the coordinate $\langle z\rangle$. Assuming a symmetric potential, i.e. $f(z)=f(-z)$, we find that $\phi^{2}(z) +\chi^{2}(z)$ is an even function. Futher we may assume that for bound states $\phi(z)$ and $\chi(z)$ are real functions. Thus we observe for the expectation value
\begin{equation}
 \left\langle z\right\rangle =\int\limits_{-\infty}^{\infty}\text{d}z\;z\left(\phi^{2}(z) +\chi^{2}(z)
\right)=0.
\end{equation}
For bound states one can assume that 
\begin{equation}
 z\left(\phi^{2}(z) +\chi^{2}(z)\right)\rightarrow 0,\; |z|\rightarrow\infty
\end{equation}
and that the normalization condition
\begin{equation}
 \int\limits_{-\infty}^{\infty}\text{d}z\;\left(\phi^{2}(z) +\chi^{2}(z)
\right)=1<\infty
\end{equation}
holds. Integration by parts yields
\begin{equation}
\int\limits_{-\infty}^{\infty}\text{d}z\;z\phi\left(z\right)  \frac{d\phi\left(z\right)}{dz}\,=\,-\frac{1}{2}\int\limits_{-\infty}^{\infty}\text{d}z\;\phi^{2}\left(z\right).
\end{equation}
The same holds for $\chi(z)$. Using these relations and the Dirac equations (\ref{eq:DiracDimensionless}) we find
\begin{equation}
\left|\int\limits_{-\infty}^{\infty}\text{d}z\;
z\phi(z)\chi(z)\right|
=\frac{1}{4}.
\label{eq:overlap}
\end{equation}
Applying the arithmetic mean inequality
\begin{equation}
 \phi(z)\chi(z)\leq \frac{1}{2}(\phi^{2}(z) +\chi^{2}(z))
\end{equation}
to eq. (\ref{eq:overlap}) we find
\begin{equation}\int\limits_{-\infty}^{\infty}\text{d}z\;
|z|\left(\phi^{2}(z) +\chi^{2}(z)\right)\geq\frac{1}{2}.
\label{eq:expectation1}
\end{equation}

Next we calculate the expectation value
\begin{equation}
\langle z^2\rangle\,=\, \int\limits_{-\infty}^{\infty}\text{d}z\;z^2\left(\phi^{2}(z) +\chi^{2}(z)\right).
\end{equation}
Using the Schwarz inequality on eq.(\ref{eq:expectation1}) we obtain
\begin{align}
\frac{1}{4}\leq & \left( \int\limits_{-\infty}^{\infty}\text{d}z\;
		|z|\left(\phi^2(z)+\chi^2(z)\right) \right)^2 \nonumber \\
\leq & \left(\int\limits_{-\infty}^{\infty}\text{d}z\;z^2\big(\phi^{2}(z) +\chi^{2}(z)\big) \right) \nonumber \\
 & \qquad \times \left(\int\limits_{-\infty}^{\infty}\text{d}z
\big(\phi^{2}(z) +\chi^{2}(z)\big)=1 \right) \nonumber \\
=\,& \langle z^2\rangle.
\end{align}
Equality only holds if $\chi(z)=\kappa\phi(z)$ for some constant $\kappa$, however this linear dependence of the eigensolutions of the Dirac equation never occurs and thus we have a strict inequality.

Reintroducing dimensional variables according to eq.(\ref{eq:dimensionlessQuantities}) we finally observe
\begin{equation}
 \langle\Delta z\rangle=\sqrt{\langle z^2\rangle-\langle z\rangle^2}>\frac{\lambda_C}{2}.
\end{equation}
This result coincides with the one obtained from the Heisenberg uncertainty relation.

It is worth noticing that this result not only applies to external scalar potentials but also to potentials created by nonlinear interactions, as e.g. given in the nonlinear massive Thirring model \cite{Thirring}. This can be seen by observing that any additional term entering eq.(\ref{eq:DiracDimensionless}) with the same sign, cancels when calculating eq.(\ref{eq:overlap}). Thus any solitonic solution of this model cannot become smaller than half its Compton wave length.

This might have applications when combining the relativistic Dirac-like light \cite{Otterbach-PRL-2009} with an intrinsic Kerr-type nonlinerarity \cite{Fleischhauer-PRL-2008b}. This system then provides a natural realisation of a massive Thirring model, where the relativistic particles are massive photons with an externally controlable mass and the nonlinearity is provided by interactions with an atomic ensemble.

\section{Conclusion}
In this brief report we derived a strict inequality for the lower bound on the coordinate variance $\langle \Delta z\rangle$ of a Dirac particle in a 1D scalar and symmetric external potential. We found that the result coincides with the estimate obtained using the Heisenberg uncertainty relation in conjunction with the nonrelativistic expression for momentum $\langle \Delta z\rangle>\lambda_C/2$. I.e. any localization of a relativistic Dirac particle has to be larger than half the corresponding Compton length.
Finally we noted that this result is also applicable to more general potentials generated e.g. by nonlinear interactions, as in the Thirring model \cite{Thirring}.

The financial support of the DFG through the GRK 792 and of the Research Center OPTIMAS is gratefully acknowledged.

\end{document}